# Electron Beam Formation and its Effect in Novel Plasma-optical Device for Evaporation of Micro-droplets in Cathode ARC Plasma Coating

*A.A. Goncharov[a], V.I. Maslov[b*], L.V. Naiko[a]*

[a] *Institute of Physics NASU, Kiev, Ukraine*

[b] *NSC Kharkov Institute of Physics and Technology, Kharkov, Ukraine*

[*] *vmaslov@kipt.kharkov.ua*

The additional pumping of energy into arc plasma flow by the self-consistently formed radially directed beam of high-energy electrons for evaporation of micro-droplets is considered. The radial beam appears near the inner cylindrical surface by secondary ion - electron emission at this surface bombardment by peripheral arc plasma flow ions. The beam is accelerated by electric potential jump, appeared in a cylindrical channel of the plasma-optical system in crossed radial electrical and longitudinal magnetic fields. The high-energy electrons pump, during the time of micro-droplet movement through the system, the energy, which is sufficient for evaporation of micro-droplets.

## 1. INTRODUCTION

Vacuum-plasma technologies of film deposition are now applied extensively [1]. Vacuum-arc method provides excellent adhesion of coatings to the substrate surface and high deposition rates. Cathode sputtering of vacuum arc by cathode spot determines the generation of flows of ions and micro-droplets. The micro-droplet component of erosion for the majority of metals is a significant part of a general loss of the cathode material in an arc, comparable with ion component. The micro-droplets in the formed ion plasma flow restrict the applicability of this method of film synthesis. Therefore for prevention of loss of substance and for prevention of influence of micro-droplets on a substrate it is necessary to eliminate micro-droplets. In [2] the conclusion has been formulated that micro-droplets cannot be evaporated in the arc plasma flow without additional energy source. Firstly attempt of micro-droplet evaporation without their removal was performed in [3]. In [4] it has been shown that in a discharge with hollow cathode the regime may exist with the double electric layer at the cathode and with high current density of fast electrons in the discharge volume, which is significantly larger than current density from cathode. Similar problems have been investigated in [5-10]. Experiments [3] demonstrate the high efficiency of the novel plasma-optical system for evaporation of micro-droplets. The additional pumping of energy into arc plasma flow by the self-consistently formed radial beam of high-energy electrons for evaporation of micro-droplets is considered in the paper. The beam is formed by double layer, appeared in a cylindrical channel of the novel plasma-optical system in crossed radial electrical and longitudinal magnetic fields. High-energy electrons appear near the inner cylindrical surface by secondary ion - electron emission at this surface bombardment by peripheral flow ions.

## 2. DYNAMIC OF ARC PLASMA FLOW IN PLASMA-OPTICAL SYSTEM

An effective system of additional energy pumping into the unit by the self-consistent formed radial beam of high-energy electrons to avaporize micro-droplets in the dense arc plasma flow is investigated in this paper. Plasma source is in a plasma-optical system at its beginning. The plasma flow is propagated from the source. The anode of arc discharge is also diaphragm. Micro-droplets fly as cone. The flow passes through the hole, part of the micro-droplets is on hole wall. After hole the flow is propagated in the system. The system is cylinder $C_1$ of length L and diameter D, on which the voltage U is applied. This leads to the formation in the vicinity of the cylinder layer of thickness $\Delta$ with a large electric field $E_r$. The width of the layer $\Delta$ is small, $\Delta \ll D$ [3]. Thus electrical potential $\varphi_0$ is concentrated in jump.

The micro-droplet velocity equals $V_d=10 \div 100$ m/s. The average micro-droplet size is equal to 1µm.

The system is in a magnetic field $H_0$ of short coil. So the following inequalities are correct.

$$\rho_{He} \ll D \ll \rho_{Hi}. \quad (1)$$

$\rho_{He}$ and $\rho_{Hi}$ are the electron and ion Larmour radiuses. Electrons are magnetized and ions are not magnetized. The electron mobility across a magnetic field is strongly suppressed. The electron movement along magnetic field is free up to region of electric potential jump. Under these conditions the magnetic field lines are equipotential up to region of electric potential jump. I.e. the magnetic field lines are equipotential inside flow. Then in space, filled with plasma, the electrical field is created, the form of which is approximately similar the structure of magnetic field lines. Because the electrons of the flow are magnetized, they in the field of the short coil are displaced to its axis, damping expansion of the flow due to electric field $E_p$ of plasma flow polarization. Thus, with increase of $H_0$ the near axis density of flow increases. I.e. $H_0$ controls the flow electrons, and they in turn keep ions from radial expansion due to $E_p$. $E_p$ can be estimated by $E_p = 2\varepsilon_i/eR_{ic}$. $R_{ic}$ is the radius of flow curvature; $\varepsilon_i$ is the ion energy.

$E_r$ in the layer $\Delta$ accelerates ions to an internal surface of the cylinder. The voltage U=1÷3 kV is supplied on cylindrical electrodes. As a result of secondary ion - electron emission the beam of electrons with velocity $V_b \approx (2e\varphi_0/m_e)^{1/2}$ and current density ($j_b=\gamma j_{is}$, $\gamma$ is the rate of secondary ion-electron emission) is formed in a thin layer

$\Delta \ll V_b/\omega_{He} = \rho_e = V_r|_{r=D/2}/\omega_{He}$ ($\omega_{He}$ is the electron cyclotron frequency). This beam, from the lateral surface to the axis (see Fig. 1), is an additional effective source of energy for evaporation of micro-droplets. During time of micro-droplet motion through system $L/V_d$ with velocity $V_d = 10 \div 100$ m/s the beam pumps large energy $\Delta\varepsilon_b$ into system.

Let us show that the collision frequency $\nu_{be}$ of high-energy electrons with plasma electrons satisfies inequality

$$\nu_{be} \ll V_b/D. \qquad (2)$$

Really we have

$$\nu_{be} \approx 7.7 \times 10^{-6} \varepsilon_b^{-3/2} n_e \lambda_{ee}.$$

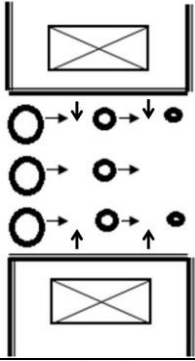

*Fig. 1. The scheme of system with self-consistently formed electron beam for evaporation of micro-droplets in arc plasma flow in vacuum-arc deposition technology*

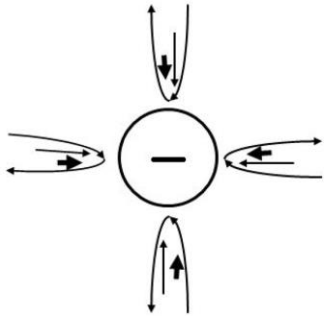

*Fig. 2. Streams of high-energy electrons, also plasma electrons and ions to micro-droplet*

For $\varepsilon_b = 1$ keV, $\lambda_{ee} \approx 10$, $n_e = 10^{12}$ cm$^{-3}$ one can derive $\nu_{be} \approx 2.6 \times 10^3$ s$^{-1}$, $V_b/D \approx 4.4 \times 10^8$ s$^{-1}$ for $D = 7$ cm. One can see that high-energy electrons many times cross the dense arc plasma flow during free movement, without taking into account the collective processes. Thus, high-energy electrons are accumulated in volume and slowly transfer energy for the micro-droplet evaporation.

One can find that dimension $R_b = \pi V_b/\omega_{He}$ of radial

$$R_b = \pi\sqrt{2\varepsilon_b m_e}\,(c/eH_0) = 0.84 \text{ cm}$$

oscillations of high-energy electrons for $\varepsilon_b = 1$ keV, $H_0 = 400$ Oe is smaller in comparison with the system radius $D/2$. High-energy electrons cross the cross-section of arc plasma flow due to their collisional mobility. It provides evaporation of micro-droplets throughout the cross-section of the arc plasma flow.

One can see that $\gamma_b \approx V_b/D \approx 3.3 \cdot 10^8$ c$^{-1}$ the rate of development of beam-plasma instability (BPI) $\gamma_b$ is close to reverse time $V_b/D$ of crossing the cross-section of flow

$$\gamma_b = \omega_{pe}(n_b/n_e)^{1/3} \approx 2.9 \times 10^8 \text{ c}^{-1}. \qquad (3)$$

However, the propagation of the beam to the axis the beam density increases and the condition of beam-plasma instability is broken. But even if for one intersection beam-plasma instability does not develop high-energy electrons are accumulated in the form of two opposing beams, $n_b$ increases and instability is developed faster. I.e. beam-plasma instability may be a mechanism for converting the beam energy into a flow.

We define, whom mostly high-energy electrons transfer energy due to collisions: drops, electrons or ions of flow. For that we compare the collision frequency of high-energy electrons with micro-droplets $\nu_{bd}$, electrons $\nu_{be}$ and ions $\nu_{bi}$ of plasma flow. We use drop radius $r_d = 1\mu$m. We derive the expression for micro-droplet density $n_d$ using what weight of drops is compared with the weight of ion component

$$n_i \approx n_d(4\pi r_d^3/3)n_{ss}. \qquad (4)$$

$n_i = 10^{12}$ cm$^{-3}$ is the ion density, $n_{ss}$ is the solid state density. Using it one can derive

$$n_d \approx 2.5 \text{ cm}^{-3}, \quad \nu_{bd} = 4\pi r_d^2 n_d V_b = 0.69 \times 10^3 \text{ s}^{-1}. \qquad (5)$$

One can see that $\nu_{bd}$ is less than the reverse time of micro-droplet movement ($V_d = 100$ m/s) through the system of length $L = 8.5$ cm

$$V_d/L \approx 10^3 \text{ s}^{-1}. \qquad (6)$$

For $n_e = n_i = 10^{12}$ cm$^{-3}$ and $V_b = (2\varepsilon_b/m_e)^{1/2}$, $\varepsilon_b = 1$ keV one can see that

$$\nu_{be} \approx 2.5 \times 10^3 \text{ s}^{-1}, \quad \nu_{bi} \approx 1.25 \times 10^3 \text{ s}^{-1}. \qquad (7)$$

$\nu_{bd} < \nu_{be}$, $\nu_{bd} < \nu_{bi}$. But growth rate of beam-plasma instability is larger than $\gamma_b \gg \nu_{bd}, \nu_{be}, \nu_{bi}$. Consequently, for considered parameters high-energy electrons pump energy into plasma electrons through beam-plasma instability. Plasma electrons transfer energy to micro-droplets directly and through ions, accelerating them to micro-droplet.

Let us estimate the number of ions $N_{id}$, which bombard micro-droplet during the time of motion of micro-droplet through chamber, in comparison with the number of atoms in the micro-droplet $N_{ad}$.

$$N_{id} \approx \pi r_d^2 n_i \sqrt{2\varepsilon_i/m_i}\,(L/V_d) = 0.96 \times 10^{10}, \quad r_d = 1\mu\text{m}, \qquad (8)$$

$$N_{ad} = \pi(r_d^3/3)n_{ss} \approx r_d^3 n_{ss} = 10^{11}.$$

One can see that the number of bombarding ions is not enough to vaporize micro-droplet.

Now we estimate the required flow of ions from the plasma to cylindrical electrodes for formation of sufficient high-energy electrons. Let us compare ion stream density $j_{is}$ on the inner cylindrical surface for beam formation with density of all ion stream $j_i$.

$$I_{is} = 2\pi RL j_{is} \text{ та } I_i = \pi R^2 j_i. \qquad (9)$$

The area of the cylindrical surface equals

$$S_w = 2\pi RL. \qquad (10)$$

For application it is necessary that ration of these flows should be small

$$(j_i/j_{is})(R/2L) \geq 1, \qquad (11)$$

because ion stream should not be lost strongly. If $R = 3.5$ cm and $L = 8.5$ cm, then

$$(j_{is}/j_i) \leq 20\%. \qquad (12)$$

One can control: what part of $j_{is}$ is supported by ions of dense arc plasma stream and what part of $j_{is}$ is supported by ions, newly formed by micro-droplet evaporation and ionization. One can provide that all ions of $j_{is}$ are newly formed ions.

Let's estimate $\Delta$. For that, equating the two currents near the inner cylindrical surface

$$j_B = 0.4(T_e/M_i)^{1/2}n_0 = j_i = (4/9\pi)(1/2em_i)^{1/2}\varphi_0^{3/2}/\Delta^2,$$

$$\Delta/r_{de} = (e\varphi_0/T_e)^{3/4} 4/3(0.4\sqrt{2})^{1/2} \gg 1.$$

One can see that $\Delta$ is large in comparison with Debye radius of electrons $r_{de}$.

One can show that $\Delta$ is smaller than the dimension of the radial oscillations of high-energy electrons $\rho_e = eE_r/m_e\omega_{He}^2$

$$\Delta \ll \rho_e = eE_r/m_e\omega_{He}^2.$$

If about 50% of weight in drops, the number of atoms, which are needed to evaporate, is approximately equal to the number of flow ions $N_a \approx n_i \pi L(D/2)^2$. We use $\varepsilon_b = 1$ keV and the configuration of electrodes such that 20% of the flow ions creates high-energy electrons. The coefficient of secondary ion-electron emission is about 0.1. Then $\Delta\varepsilon_b = 5.9$ keV are pumped by high-energy electrons during the time $L/V_d$ of drop movement through the system and it is attributed to a single atom, which is to be evaporated. This is sufficient for evaporation of micro-droplets.

The collision frequency $v_{ed}$ of micro-droplets with flow electrons is more than with high-energy electrons $v_{bd}$, because

$$v_{bd} = \pi r_d^2 j_b \ll v_{ed} = \pi r_d^2 n_e V_{the} \exp(-e\varphi_d/T_e), \quad (13)$$

$$j_b = 0.4\gamma n_i V_s \ll n_i V_{0i}/4.$$

But the energy, which is pumped into the micro-droplet by high-energy electrons, is larger than by flow electrons

$$v_{bd}\varepsilon_b/v_{ed}T_e = 0.16(T_e/2\varepsilon_i)^{1/2}(\varepsilon_b/T_e) \approx 10. \quad (14)$$

$$4n_e V_{the} \exp(-e\varphi_d/T_e) \approx n_i V_{0i}, \quad V_{0i} = \sqrt{2\varepsilon_i/m_i}. \quad (15)$$

$\varepsilon_i$ is the energy of ions of arc plasma flow.

Regime with strongly charged micro-droplet, i.e. micro-droplet with a large electric potential $\varphi_d$, is disadvantageous for the evaporation of micro-droplets. High-energy electrons can effectively ensure the reduction of micro-droplet potential. If $e\varphi_d \gg T_e$, micro-droplets are evaporated due to ion bombardment, but in this case the energy flow is not large. Indeed, the full flow of energy $\varepsilon_\Sigma$ to micro-droplet approximately equals

$$\varepsilon_\Sigma \approx \pi r_d^2 \begin{bmatrix} 4n_e V_{the} T_e \exp(-e\varphi_d/T_e) + \\ +1,6\gamma n_i V_s \varepsilon_b + n_i V_{0i}(\varepsilon_i + e\varphi_d) \end{bmatrix}. \quad (16)$$

1-st member is the effect of plasma electrons, 2-nd member is the contribution of high-energy electrons, 3-rd member is the contribution of flow ions. At steady state, the balance of flows of plasma electrons and ions in micro-droplet is approximately equal to

$$4n_e V_{the} \exp(-e\varphi_d/T_e) \approx n_i V_{0i}.$$

Then the ratio of energy flows of plasma electrons $j_{e\varepsilon}$ and ions $j_{i\varepsilon}$ to micro-droplet equals

$$j_{e\varepsilon}/j_{i\varepsilon} = T_e/(\varepsilon_i + e\varphi_d).$$

One can see that in absence of high-energy electrons at $e\varphi_d \gg T_e$ energy flow is determined by the ions. However, it does not exceed $j_{b\varepsilon}$ the energy flow of high-energy electrons

$$j_{b\varepsilon}/j_{i\varepsilon} = 0.4\gamma n_i V_s \varepsilon_b/n_i V_{0i}(\varepsilon_i + e\varphi_d) = $$
$$= 0.04(T_e/2\varepsilon_i)^{1/2}[\varepsilon_b/(\varepsilon_i + e\varphi_d)]$$

It is useful to reduce $\varphi_d$ because at its reducing, although the ion energy flow decreases to a micro-droplet, but the energy flow of plasma electrons to micro-droplet is increased even faster. Indeed, the energy flow of plasma electrons to micro-droplet at $\varphi_d = 0$ is larger than the ion energy flow to a micro-droplet at $e\varphi_d \gg T_e$, because following inequality is correct

$$4n_{0e} V_{the} T_e/n_i V_{0i}(\varepsilon_i + e\varphi_d) = $$
$$= 4(T_e/2\varepsilon_i)^{1/2}(m_i/m_e)^{1/2} T_e/(\varepsilon_i + e\varphi_d) > 1.$$

This reduction of $\varphi_d$ is carried out automatically in nonequilibrium arc plasma flow and by high-energy electrons. In particular, high-energy flow ions lead not only to some evaporation of micro-droplets in their motion to the substrate, but also to reduce of the micro-droplet electric potential

$$\varphi_d = (T_e/e)\ln\left[4(m_i/m_e)^{1/2}(T_e/2\varepsilon_i)^{1/2}\right], \quad (17)$$

facilitating the energy exchange of plasma electrons with micro-droplets. Due to $\varphi_d \neq 0$ many plasma electrons do not reach the surface of the micro-droplet, high-energy electrons also lose some energy in $\varphi_d$.

Energy of high-energy electrons spent on: inelastic collisions with plasma electrons, inelastic collisions with plasma ions, heating with plasma electrons and ions of evaporated layer to a plasma temperature, ionization of evaporated atoms, inelastic collisions with micro-droplets, energy loss on the beam-plasma instability development.

The energy, which is pumped in micro-droplets, spent on: heating, evaporation and dispersion; on thermal radiation; on electrons of the secondary emission; on electrons of the thermal emission. Thus, one can write the equation of balance of energy flows on the micro-droplet surface

$$Q_\Sigma = \frac{L}{V_{dr}}\begin{bmatrix} I_e(r_d)(T_e - e\varphi_d) - gI_b(r_d)e\varphi_d + \\ +I_i(r_d)(\varepsilon_{ii} + \varepsilon_v + \varepsilon_i + e\varphi_d) + I_b(r_d)(\varepsilon_b - e\varphi_d) \end{bmatrix}$$
$$- \frac{L}{V_{dr}}\left[gI_b(r_d)e\varphi_d + I_{th}(T_d, r_d)e\varphi_d + \alpha\sigma T_d^4\right], \quad (18)$$

$$Q_\Sigma = Q_1 + Q_2 + Q_3.$$

$Q_1 = cm_{dr}(T_2 - T_1)$ is the energy for micro-droplet heating from $T_1$ to evaporation temperature $T_2$; $m_{dr}$ is the micro-droplet weight, c is the unit heat capacity; $Q_2 = r\delta m_{dr}$ is the energy for vaporization of the part $\delta m_{dr}$ of the micro-droplet, which had time to evaporate, r is the unit heat of vaporization; $Q_3$ is the energy for spraying of part of micro-droplet atoms by ions. $I_e(r_{dr})$, $I_i(r_{dr})$, $I_b(r_{dr})$ are the currents of flow electrons, ions and high-energy electrons to micro-droplet. $\varepsilon_v$ is the condensation energy; $\varepsilon_{ii}$ is the ionization energy of flow ions; $T_{dr}$ is the micro-droplet temperature; $\alpha$ is the emissivity; $\sigma$ is the Stefan-Boltzmann constant; g is the constant of secondary electron-electron emission; $I_{th}(T_{dr}, r_{dr})$ is the current of thermionic emission.

## 3. CONCLUSION

Method of formation of jump of electrical potential near the inner surface of the cylinder and of the injection of self-consistent electron beam across the flow with micro-droplets has been proposed for efficient evaporation of micro-droplets in the arc plasma flow.

**ФОРМИРОВАНИЕ ЭЛЕКТРОННОГО ПУЧКА И ЕГО РОЛЬ В НОВОЙ ПЛАЗМО-ОПТИЧЕСКОЙ СИСТЕМЕ ДЛЯ ИСПАРЕНИЯ КАПЕЛЬ В ДУГОВОЙ ПЛАЗМЕ**

*А.А. Гончаров, В.И. Маслов, Л.В.Найко*

Рассматривается дополнительная накачка энергии в поток дуговой плазмы с помощью самосогласованно образуемого радиального пучка электронов для испарения капель. Пучок появляется вблизи внутренней цилиндрической поверхности за счет вторичной ионно - электронной эмиссии при ее бомбардировке периферийными ионами потока. Пучок ускоряется скачком электрического потенциала, который появляется в цилиндрическом канале плазмо-оптической системы в скрещенных полях. Высокоэнергетичные электроны накачивают за время движения микро-капель через систему энергию, которая достаточна для испарения микрокапель.

**ФОРМУВАННЯ ЕЛЕКТРОННОГО ПУЧКА І ЙОГО РОЛЬ В НОВІЙ ПЛАЗМО-ОПТИЧНІЙ СИСТЕМІ ДЛЯ ВИПАРОВУВАННЯ КРАПЕЛЬ У ДУГОВІЙ ПЛАЗМІ**

*О.А.Гончаров, В.І.Маслов, Л.В.Найко*

Розглядається додаткове накачування енергії в потік дугової плазми за допомогою самоузгоджене утвореного радіального пучка електронів для випаровування крапель. Пучок з'являється поблизу внутрішньої циліндричної поверхні за рахунок вторинної іонно - електронної емісії при її бомбардуванні периферійними іонами потоку. Пучок прискорюється стрибком електричного потенціалу, який з'являється в циліндричному каналі плазмо-оптичної системи в схрещених полях. Високоенергетичні електрони накачують за час руху крапель через систему енергію, яка достатня для випаровування крапель.